\documentclass[amsmath,amsfonts,amssymb,superscriptaddress,preprint]{revtex4}
\usepackage{graphicx}
\usepackage{amsmath}
\usepackage{amsfonts}
\usepackage{amssymb}

\begin{document}

\title{Snowmass 2001: Jet Energy Flow Project\footnote{Contribution 
to the P5 Working Group on QCD and Strong Interactions at
Snowmass 2001}}

\author{C.F. Berger}
\affiliation{State University of New York, Stony Brook}

\author{E.L. Berger}
\affiliation{Argonne National Laboratory}

\author{P.C. Bhat}
\affiliation{Fermi National Accelerator Laboratory}

\author{J.M. Butterworth}
\affiliation{University College London}

\author{S.D. Ellis}
\affiliation{University of Washington}

\author{B. Flaugher}
\affiliation{Fermi National Accelerator Laboratory}

\author{W.T. Giele}
\affiliation{Fermi National Accelerator Laboratory}

\author{W. Kilgore}
\affiliation{Brookhaven National Laboratory}

\author{A. Kulesza}
\affiliation{Brookhaven National Laboratory}

\author{S. Lammers}
\affiliation{University of Wisconsin/DESY}

\author{S. Magill}
\affiliation{Argonne National Laboratory}

\author{H. Prosper}
\affiliation{Florida State University}

\date{February 19, 2002}

\begin{abstract}
Conventional cone jet algorithms arose from heuristic considerations of LO
hard scattering coupled to independent showering. \ These algorithms
implicitly assume that the final states of individual events can be mapped
onto a unique set of jets that are in turn associated with a unique set of
underlying hard scattering partons. \ Thus each final state hadron is assigned
to a unique underlying parton. \ The Jet Energy Flow (JEF)\ analysis described
here does not make such assumptions. \ The final states of individual events
are instead described in terms of flow distributions of hadronic energy.
\ Quantities of physical interest are constructed from the energy flow
distribution summed over all events. \ The resulting analysis is less
sensitive to higher order perturbative corrections and the impact of showering
and hadronization than the standard cone algorithms.
\end{abstract}

\preprint{UW/PT-02-03}

\maketitle

%\homepage[]{Your web page}
%\thanks{}
%\altaffiliation{}
%

%insert suggested PACS numbers in braces on next line
% \pacs{}
%

%\maketitle must follow title, authors, abstract and \pacs
%

\subsection{Introduction}

The goal of a jet algorithm is to provide a precise mapping between the
observed, long distance hadronic final states in high energy interactions and
the underlying energetic partons participating in the true short-distance,
hard scattering process\cite{workshops}. \ To appreciate this goal imagine
comparing the observed final states detected by ``real'' detectors that have
sizes of order centimeters to meters to what would be observed with a detector
whose size is characterized by a distance scale of a fraction of a fermi.
\ Between these two scales, \textit{i.e}., from a fraction of a fermi to
centimeters, the short-distance state evolves via higher order perturbative
processes and the physics associated with showering and hadronization. \ These
processes allow the short-distance partons, along with the spectators, to
evolve into the observed hadrons. \ During the evolution the 4-momentum
associated with an initial short-distance parton is spread out over a number
of final state hadrons occupying an extended region of phase space. \ To
achieve our stated goal, the jet algorithm tries to identify these ``related''
hadrons into a single jet, whose total 4-momentum should track that of the
initial parton. \ To perform this task with precision it is important that the
results of applying the jet algorithm are insensitive to both higher order
corrections and fluctuations in the showering/hadronization processes. \ The
results of the jet algorithm should also be insensitive to the smearing
effects of the detection process itself, \textit{e.g}., due to the granularity
of the detector.

Current jet algorithms attempt to achieve the stated goal in a quite singular
way by assigning the observed hadrons to \emph{unique} jets on an
\emph{event-by-event} basis. \ This identification proceeds in the face of the
fundamental fact that such a unique assignment cannot be precise. \ While the
underlying parton that initiates the jet is a nonsinglet under the color
symmetry of QCD, the hadrons are all color singlets. \ At the very least, the
jets in the final state must represent the correlated evolution of more than
one short-distance parton or of a short-distance parton correlated with a
spectator parton, \textit{i.e}., a color singlet set of initial partons. \ In
some sense the most extreme approach is represented by cone algorithms as
advocated at the 1990 Snowmass Workshop\cite{Snowmass}. \ Cone algorithms
associate hadrons into jets by identifying those that are nearby in angle.
\ The underlying assumption is that the extra radiation produced by higher
order corrections, showering and hadronization around an energetic parton
appears symmetrically, \textit{i.e}., occurs independently of the other color
charged objects in the final state. \ The alternative jet algorithm, the
$K_{T}$ algorithm\cite{ktcluster}, uses nearness in momentum space to identify
the members of a jet and, at least to some extent, recognizes the
``color-connectedness'' of the radiation producing the final state. \ However,
as noted above, both of these algorithms associate individual hadrons with
unique jets on an event-by-event basis. \ We know that this procedure is only
an approximation (due to the color conservation issue) and can lead to
undesirable dependence (at least at the 10\% level) on the details of the
showering and hadronization processes. \ In the context of cone jet algorithms
this latter point is discussed in more detail in another contribution to these
Proceedings\cite{better jets}.

The Jet Energy Flow (JEF) approach described in this note is a simplified
version of the more completely developed C-continuous observables or
C-algebra formalism of F. V.
Tkachov\cite{Tkachov} for describing energy flow in hadronic
collisions\cite{Eflow}. \ JEF accepts the reality that the hadronic final
state represents the collective radiation from several out-flowing color
charges, \textit{i.e}., the underlying short-distance partons. \ No attempt is
made to associate individual hadrons with unique jets, \textit{i.e}., with
unique underlying partons, on an event-by-event basis. \ Yet the energy flow
pattern of an event still provides a footprint of the underlying partons, from
which much of the same information provided by the standard jet algorithms can
be extracted. \ As the subsequent discussion will indicate, it is a more
reliable characterization of the event in the sense of exhibiting a reduced
sensitivity to the showering and hadronization processes. \ The challenge in
the JEF type analysis is to define observables that offer an informative
comparison between theory and experiment.

\subsection{The JEF Formalism}

A primary strength of the JEF approach is that, in contrast with the usual
algorithmic approach to jet identification, the JEF formalism generates,
event-by-event, a smooth distribution to characterize each event. In that
sense, the JEF formalism is more analytic. For example, in the application of
the cone algorithm the goal of identifying unique jets leads to the
``stability'' constraint\cite{better jets}. \ A set of hadrons or partons that
lie within a cone of a defined size $R$ are identified as constituting a jet
if and only if the energy-weighted centroid of the set of particles coincides
with the geometric center of the cone. \ This constraint results in the
non-analytic structure of the implementation of the algorithm, typically in
the form of step functions with complicated arguments. \ Only limited (and
typically complicated) regions of the multi-particle phase space contribute to
a jet. \ No such constraint arises in the JEF\ analysis. This distinction has
several important consequences.

\begin{enumerate}
\item  The more inclusive and analytic calculations characteristic of a JEF
analysis are more amenable to resummation techniques and power corrections
analysis in perturbative calculations.

\item  Since the required multi-particle phase space integrations are largely
unconstrained, \textit{i.e.,} more analytic, they are easier (and faster) to
implement. \ Programs like JETRAD spend considerable computer time simulating
the complicated phase space required by the algorithmic style jet algorithms.

\item  Since the analysis does not identify jets event-by-event, the analysis
of the experimental data from an individual event should proceed more quickly.

\item  Signal to background optimization can now include the JEF\ parameters
(and distributions). \ One cannot typically optimize a standard jet algorithm
except for a limited number of parameters.
\end{enumerate}

\begin{figure}
[ptb]
\begin{center}
\includegraphics[
height=3.2517in,
width=3.4411in
]%
{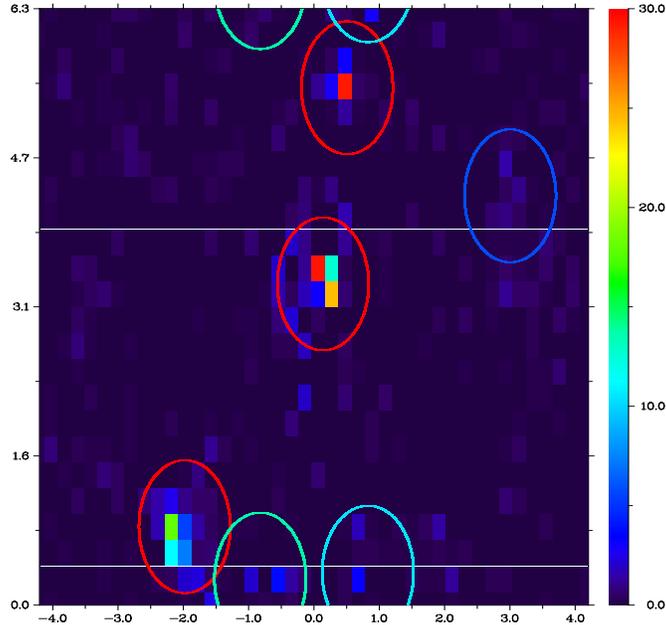}%
\caption{Transverse energy (in GeV color coding) calorimeter readout in a
typical CDF event as a function of pseudorapidity and azimuthal angle. \ The
reconstructed CDF cone jets are indicated by the circles (with the same GeV
color coding). }%
\label{calread}%
\end{center}
\end{figure}

We can define the fundamental distribution of the JEF analysis as follows.
\ We start with a set of 4-vectors, $p_{\mu}=\left(  E,p_{x},p_{y}%
,p_{z}\right)  $, that represent either the partons in a perturbative
calculation or the hadrons in a simulated or real event. \ In the latter case
these 4-vectors might correspond as well to the location and energy deposited
in individual calorimeter cells. \ If a given event corresponds to the
measurement of $N$ such 4-vectors, $\left\{  p_{\mu}^{i}\right\}  _{i=1}%
^{N}=\left\{  \left(  E^{i},\overrightarrow{P^{i}}\right)  \right\}
_{i=1}^{N}$, we have the 4-vector distribution for that event defined by
\begin{equation}
P_{\mu}\left(  \widehat{P}\right)  =\sum\limits_{i=1}^{N}p_{\mu}^{i}%
\delta\left(  \widehat{P}-\widehat{P^{i}}\right)  , \label{4-vector}%
\end{equation}
where the directional unit vector is defined by $\widehat{P}\left(  m\right)
=\overrightarrow{P}/\left|  \overrightarrow{P}\right|  $ with the
2-dimensional angular variable defined as $m=\left(  \theta,\phi\right)  $
(typical of lepton colliders) or $m=\left(  \eta,\phi\right)  $ (typical of
hadron colliders, where $\eta$ is the pseudorapidity, $\eta=\ln\left(
\cot\theta/2\right)  $). \ This expression defines the underlying energy flow
via $E\left(  m\right)  =P_{0}\left(  m\right)  $ with a corresponding
expression for the underlying longitudinal momentum flow $P_{z}\left(
m\right)  $. \ For the case of hadronic colliders the more familiar underlying
transverse energy flow is defined by the composite quantity $E_{T}\left(
m\right)  =\sqrt{P_{x}^{2}\left(  m\right)  +P_{y}^{2}\left(  m\right)  }$ or
$E_{T}\left(  m\right)  =E\left(  m\right)  \times\sin\theta\left(  m\right)
$. \ Clearly many quantities can be constructed from the 4-momentum
distribution of Eq. \ref{4-vector}, including the usual cone jet algorithm.
\ The $E_{T}\left(  \eta,\phi\right)  $ distribution for a typical CDF jet
event is illustrated in Figure \ref{calread} along with the cone jets
``found'' with the CDF cone jet algorithm.%
%TCIMACRO{\FRAME{ftbpFU}{3.4411in}{3.2517in}{0pt}{\Qcb{Transverse energy (in
%GeV color coding) calorimeter readout in a typical CDF event as a function of
%pseudorapidity and azimuthal angle. \ The reconstructed CDF cone jets are
%indicated by the circles (with the same GeV color coding). }}{\Qlb{calread}%
%}{calread.eps}{\special{ language "Scientific Word";  type "GRAPHIC";
%display "USEDEF";  valid_file "F";  width 3.4411in;  height 3.2517in;
%depth 0pt;  original-width 6.1505in;  original-height 6.8779in;
%cropleft "0";  croptop "1";  cropright "1";  cropbottom "0";
%filename 'P5_ellis_0707_fig1.eps';file-properties "XNPEU";}}}%
%BeginExpansion
%EndExpansion

\begin{figure}
[ptb]
\begin{center}
\includegraphics[
height=3.1912in,
width=2.8617in
]%
{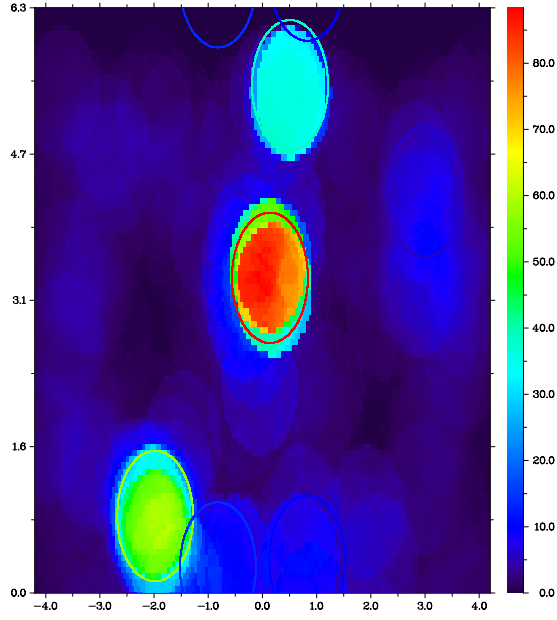}%
\caption{The $E_{T}\left(  \eta,\phi\right)  $ flow (including the factor $\pi
R^{2}$) using the CDF event of Figure 1.}%
\label{etflow}%
\end{center}
\end{figure}

\begin{figure}
[ptbptb]
\begin{center}
\includegraphics[
height=3.2578in,
width=2.9032in
]%
{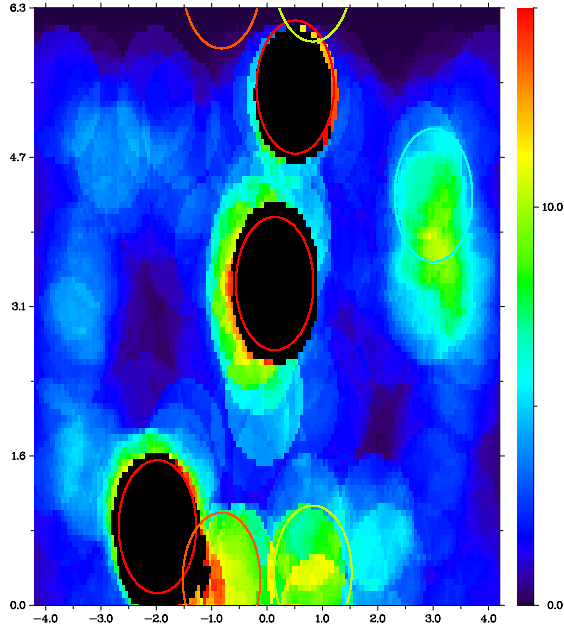}%
\caption{Same as Figure 2 except that the maximum $E_{T}$ for the color coding
is 15 GeV.}%
\label{etflow_15}%
\end{center}
\end{figure}

Using the underlying 4-vector distribution we define the jet energy flow (JEF)
via a smearing or averaging function $A$\ as%
\begin{equation}
J_{\mu}\left(  m\right)  \equiv\int dm^{\prime}\ P_{\mu}\left(  m^{\prime
}\right)  \ \times\ A\left(  m^{\prime}-m\right)  , \label{JEF}%
\end{equation}
where $A$ is normalized as%
\begin{equation}
\int dm\ A\left(  m\right)  =1. \label{norm}%
\end{equation}
A simple (but not unique) form for the averaging function in\ terms of the
general 2-tuple of angular variables $m=\left(  \alpha,\beta\right)  $, which
provides a direct comparison with the jet cone algorithm, is%
\begin{equation}
A\left(  m\right)  =A\left(  \alpha,\beta\right)  =\frac{\Theta\left(
R-r\left(  \alpha,\beta\right)  \right)  }{\pi R^{2}}=\frac{\Theta\left(
R-\sqrt{\alpha^{2}+\beta^{2}}\right)  }{\pi R^{2}}, \label{averaging}%
\end{equation}
where $R$ is the cone size and $r\left(  \alpha,\beta\right)  $ is the
distance measure in the space defined by $\left(  \alpha,\beta\right)  $.
\ For comparison with the existing jet cone analyses we will discuss the case
$m=\left(  \eta,\phi\right)  $.\ \ As a specific example we exhibit the jet
transverse energy flow (transverse JEF)%
\begin{equation}
J_{T}\left(  m\right)  =\int dm^{\prime}\ E_{T}\left(  m^{\prime}\right)
\ \times\ A\left(  m^{\prime}-m\right)  =\int dm^{\prime}\ \sqrt{P_{x}%
^{2}\left(  m^{\prime}\right)  +P_{y}^{2}\left(  m^{\prime}\right)  }%
\ \times\ A\left(  m^{\prime}-m\right)  \label{JetET}%
\end{equation}
times a factor of $\pi R^{2}$ ($E_{T}=\pi R^{2}\times J_{T})$ for the case of
$R=0.7$ in Figure \ref{etflow} and \ref{etflow_15}\ for the same event
displayed in Figure \ref{calread}. \ Clearly the same general structure is
present in all three figures.
%TCIMACRO{\FRAME{ftbpFU}{2.8617in}{3.1912in}{0pt}{\Qcb{The $E_{T}\left(
%\eta,\phi\right)  $ flow (including the factor $\pi R^{2}$) using the CDF
%event of Figure 1.}}{\Qlb{etflow}}{etflow.eps}%
%{\special{ language "Scientific Word";  type "GRAPHIC";
%maintain-aspect-ratio TRUE;  display "USEDEF";  valid_file "F";
%width 2.8617in;  height 3.1912in;  depth 0pt;  original-width 6.1661in;
%original-height 6.8779in;  cropleft "0";  croptop "1";  cropright "1";
%cropbottom "0";  filename 'P5_ellis_0707_fig2.eps';file-properties "XNPEU";}}}%
%BeginExpansion
%EndExpansion%
%TCIMACRO{\FRAME{ftbpFU}{2.9032in}{3.2578in}{0pt}{\Qcb{Same as Figure 2 except
%that the maximum $E_{T}$ for the color coding is 15 GeV.}}{\Qlb{etflow_15}%
%}{etflow_15.eps}{\special{ language "Scientific Word";  type "GRAPHIC";
%maintain-aspect-ratio TRUE;  display "USEDEF";  valid_file "F";
%width 2.9032in;  height 3.2578in;  depth 0pt;  original-width 6.122in;
%original-height 6.8779in;  cropleft "0";  croptop "1";  cropright "1";
%cropbottom "0";  filename 'P5_ellis_0707_fig3.eps';file-properties "XNPEU";}}}%
%BeginExpansion
%EndExpansion
\ Note that the transverse JEF is smeared on a scale $R$ compared to the
underlying $E_{T}$ distribution of Figure \ref{calread}. \ For comparison the
Snowmass cone jet algorithm\cite{Snowmass,EKS} identifies jets at a discrete
set of values of locations $m_{j}$ defined by the $E_{T}$ weighted cone
``stability'' constraint. \ These stable cone locations $m_{j}$ are the
solutions of the equation%
\begin{equation}
\int dm^{\prime}\ E_{T}\left(  m^{\prime}\right)  \ \times\ \left(  m^{\prime
}-m_{j}\right)  \ \times\ A\left(  m^{\prime}-m_{j}\right)  =0.
\label{stablecone}%
\end{equation}
The corresponding cone jet $E_{T}$ values are found by evaluating Eq.
\ref{JetET} (times $\pi R^{2}$) at the jet positions, $E_{T,j}=\pi R^{2}\times
J_{T}\left(  m_{j}\right)  $. \ The non-analytic character of the jet cone
algorithm referred to earlier arises from the need to solve Eq.
\ref{stablecone} and then work with only the discrete set of solutions,
\textit{i.e}.,\ the jets in an event.

\subsection{Observables}

We can now proceed to define more general observables. \ The basic assumption
of the JEF approach is that event-by-event each value of the direction $m$ is
equally likely to correspond to a jet with 4-momentum proportional to $J_{\mu
}\left(  m\right)  $. \ Relative probabilities of observables having values in
a specified range will correspond to the size of the area in $m$ covered by
the JEF with the correct range of values. \ To illustrate these ideas,
consider a general nth order observable $C_{n}$ represented by an nth order
function of $J_{\mu}\left(  m\right)  $, $C\left(  J\left(  m_{1}\right)
,\ldots,J\left(  m_{n}\right)  \right)  $. \ The corresponding event
probability distribution, including the possibility of providing a set of
angular cuts $\Omega$, is given by%
\begin{align}
P\left(  C_{n}\left|  \Omega\left(  m_{cut}\right)  \right.  \right)   &
=\left(  \prod\limits_{i=1}^{n}\int\frac{dm_{i}}{\pi R^{2}}\Omega\left(
m_{i}-m_{cut}\right)  \right)  \delta\left(  C_{n}-C\left(  J\left(
m_{1}\right)  ,\ldots,J\left(  m_{n}\right)  \right)  \right) \label{Cn}\\
&  \propto\frac{d\sigma_{\mathrm{JEF}}}{dC_{n}},\nonumber
\end{align}
where we have normalized the area to the ``cone size'' $\pi R^{2}$. \ (Note
that one might also consider applying a cut directly on the $J\left(
m\right)  $, \textit{e.g}., $J_{0}\left(  m\right)  >E_{cut}$.) \ To determine
the differential cross section for the observable $C_{n}$ from an experiment
we simply sum over events as%
\begin{equation}
\mathcal{L}\frac{d\sigma}{dC_{n}}=\sum\limits_{\mathrm{events}}P\left(
C_{n}\left|  \Omega\left(  m_{cut}\right)  \right.  \right)  , \label{diffCn}%
\end{equation}
where $\mathcal{L}$ is the integrated luminosity. \ We obtain an event
occupancy probability $O$ by integrating over the probability function%
\begin{align}
O\left(  C_{n}\left(  \min\right)  ,\left.  C_{n}\left(  \max\right)  \right|
\Omega\left(  m_{cut}\right)  \right)   &  =\int_{C_{n}\left(  \min\right)
}^{C_{n}\left(  \max\right)  }dC_{n}\ P\left(  C_{n}\left|  \Omega\left(
m_{cut}\right)  \right.  \right) \label{occupancy}\\
&  =\left(  \prod\limits_{i=1}^{n}\int\frac{dm_{i}}{\pi R^{2}}\Omega\left(
m_{i}-m_{cut}\right)  \right)  \Theta\left(  C_{n}\left(  \max\right)
-C\left(  J\left(  m_{1}\right)  ,\ldots,J\left(  m_{n}\right)  \right)
\right) \nonumber\\
&  \quad\times\Theta\left(  C\left(  J\left(  m_{1}\right)  ,\ldots,J\left(
m_{n}\right)  \right)  -C_{n}\left(  \min\right)  \right)  .\nonumber
\end{align}
This final expression indicates that we are simply calculating the relative
area in $m$ for which the JEF has the correct value to yield the desired value
of the observable. \ We can then count the number of events with effective
occupancy
number $O$ and convert it into a cross section,%
\begin{equation}
\mathcal{L}\sigma\left(  C_{n}\left(  \min\right)  ,C_{n}\left(  \max\right)
\right)  =\sum\limits_{\mathrm{events}}O\left(  C_{n}\left(  \min\right)
,\left.  C_{n}\left(  \max\right)  \right|  \Omega\left(  m_{cut}\right)
\right)  . \label{occupancysigma}%
\end{equation}
This formula can be used to construct bin values and the corresponding
distribution. \ Let us illustrate these ideas by considering some explicit examples.

\begin{enumerate}
\item  The JEF jet mass $M\left(  J\left(  m\right)  \right)  =\pi R^{2}%
\times\sqrt{J_{\mu}\left(  m\right)  J^{\mu}\left(  m\right)  }$ is an example
of a $C_{1}$ observable with a event probability distribution of the form
\begin{equation}
P\left(  M_{J}\right)  =\int\frac{dm}{\pi R^{2}}\ \delta\left(  M_{J}-M\left(
J\left(  m\right)  \right)  \right)  . \label{jetmass}%
\end{equation}
The corresponding occupancy probability has the form%
\[
O\left(  M_{J}\left(  \min\right)  ,M_{J}\left(  \max\right)  \right)
=\int\frac{dm}{\pi R^{2}}\ \Theta\left(  M_{J}\left(  \max\right)  -M\left(
J\left(  m\right)  \right)  \right)  \Theta\left(  M\left(  J\left(  m\right)
\right)  -M_{J}\left(  \min\right)  \right)
\]
Thus the fraction of the events with a JEF jet with mass in the specified
range is proportional to the fractional area in the $m$ plane occupied by
JEF\ jets with a mass value in that range. To obtain the final distribution we
sum over events. \ The (simulated) JEF jet mass distribution for the $W$ decay
into hadrons (treated as a single jet) in the process $p\overline
{p}\rightarrow H+X\rightarrow W^{+}W^{-}+X\rightarrow l\nu+\mathrm{hadrons}+X$
\ is exhibited in Figure \ref{mass}.%
%TCIMACRO{\FRAME{ftbpFU}{2.9922in}{2.8694in}{0pt}{\Qcb{The JEF jet invariant
%mass as defined in Eq. \ref{jetmass}\ for $W$ boson decay in Higgs boson
%decay.}}{\Qlb{mass}}{mass.eps}{\special{ language "Scientific Word";
%type "GRAPHIC";  maintain-aspect-ratio TRUE;  display "USEDEF";
%valid_file "F";  width 2.9922in;  height 2.8694in;  depth 0pt;
%original-width 7.8793in;  original-height 7.5567in;  cropleft "0";
%croptop "1";  cropright "1";  cropbottom "0";
%filename 'P5_ellis_0707_fig4.eps';file-properties "XNPEU";}}}%
%BeginExpansion
\begin{figure}
[ptb]
\begin{center}
\includegraphics[
height=2.8694in,
width=2.9922in
]%
{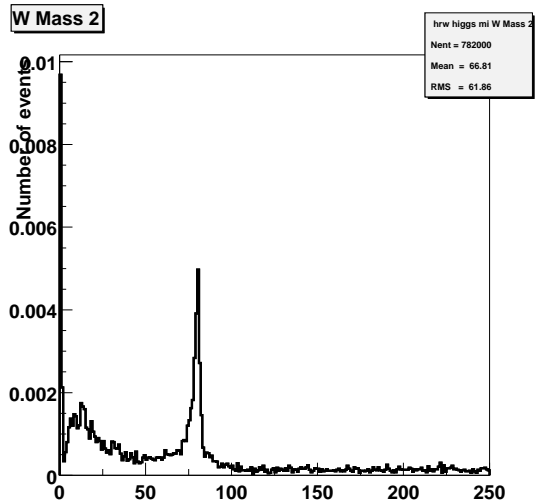}%
\caption{The JEF jet invariant mass as defined in Eq. \ref{jetmass}\ for $W$
boson decay in Higgs boson decay.}%
\label{mass}%
\end{center}
\end{figure}
%EndExpansion

\item  The JEF jet transverse energy $E_{T}$ in the variables appropriate to a
hadron collider is another $C_{1}$ observable. \ The relative
probability distribution
for a CDF type rapidity acceptance and CDF $E_{T}$ definition looks like%
\begin{equation}
P\left(  E_{T}\left|  \Omega\left(  0.1<\left|  \eta\right|  <0.7\right)
\right.  \right)  =\frac{1}{\pi R^{2}}\int_{0.1}^{0.7}d\left|  \eta\right|
\oint d\phi\delta\left(  E_{T}-E\left(  J\left(  \eta,\phi\right)  \right)
\times\sin\left(  \theta\left(  \eta\right)  \right)  \right)  \ ,
\label{probET}%
\end{equation}
where the JEF energy distribution is given by $E\left(  J\left(  \eta
,\phi\right)  \right)  =\pi R^{2}\times J_{0}\left(  \eta,\phi\right)  $. \ As
suggested above, we can obtain the effective occupancy number of JEF jets 
(per event) above an energy cut
$E_{T,\min}$ by integrating%
\begin{equation}
O\left(  E_{T,\min}\left|  \Omega\left(  0.1<\left|  \eta\right|  <0.7\right)
\right.  \right)  =\int_{E_{T,\min}}dE_{T}\ P\left(  E_{T}\left|
\Omega\left(  0.1<\left|  \eta\right|  <0.7\right)  \right.  \right)  .
\label{numjets}%
\end{equation}
These quantities as evaluated for the sample jet event of Figure 1 are
illustrated in Figure \ref{etdis}. \ The jets found by the standard CDF cone
jet algorithm are also indicated as data points in the figures and correlate
well with the peaks in the JEF\ probability distribution.
%TCIMACRO{\FRAME{ftbpFU}{5.9923in}{1.6596in}{0pt}{\Qcb{The $E_{T}$ probability
%function of Eq. \ref{probET} for the event of Figure 1: the left figure is for
%a bin width of 1 GeV and middle figure is for a bin width of 5 GeV. \ The
%right figure is the occupancy number of Eq. \ref{numjets}. \ The individual
%data points corresponding to the 5 largest energy jets found by the CDF cone
%jet algorithm and illustrated in Figure 1.}}{\Qlb{etdis}}{etdis.eps}%
%{\special{ language "Scientific Word";  type "GRAPHIC";
%maintain-aspect-ratio TRUE;  display "USEDEF";  valid_file "F";
%width 5.9923in;  height 1.6596in;  depth 0pt;  original-width 6.947in;
%original-height 1.9043in;  cropleft "0";  croptop "1";  cropright "1";
%cropbottom "0";  filename 'P5_ellis_0707_fig5.eps';file-properties "XNPEU";}}}%
%BeginExpansion
\begin{figure}
[ptb]
\begin{center}
\includegraphics[
height=1.6596in,
width=5.9923in
]%
{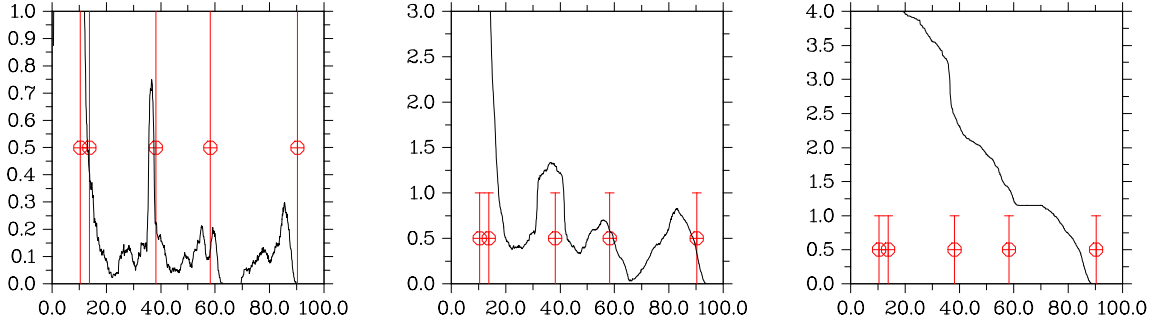}%
\caption{The $E_{T}$ probability function of Eq. \ref{probET} for the event of
Figure 1: the left figure is for a bin width of 1 GeV and middle figure is for
a bin width of 5 GeV. \ The right figure is the occupancy number of Eq.
\ref{numjets}. \ The individual data points corresponding to the 5 largest
energy jets found by the CDF cone jet algorithm and illustrated in Figure 1.}%
\label{etdis}%
\end{center}
\end{figure}
%EndExpansion

\item  The JEF di-jet invariant mass $M\left(  J,J\right)  $ is an example of
a $C_{2}$ observable with the form $M^{2}\left(  J\left(  m_{1}\right)
,J\left(  m_{2}\right)  \right)  =\left(  \pi R^{2}\right)  ^{2}\times\left(
J\left(  m_{1}\right)  +J\left(  m_{2}\right)  \right)  _{\mu}\left(  J\left(
m_{1}\right)  +J\left(  m_{2}\right)  \right)  ^{\mu}$, which assumes that the
two JEF\ jets are non-overlapping.. \ The corresponding probability
distribution is defined by%
\begin{equation}
P\left(  M_{JJ}^{2}\right)  =\iint\frac{dm_{1}dm_{2}}{\left(  \pi
R^{2}\right)  ^{2}}\delta\left(  M_{JJ}^{2}-M^{2}\left(  J\left(
m_{1}\right)  ,J\left(  m_{2}\right)  \right)  \right)  .\label{dijet}%
\end{equation}
\end{enumerate}

\subsection{An Example JEF Analysis}

As an example of a JEF style analysis we briefly review a JEF\ di-jet analysis
performed previously\cite{GandG}. \ The goal is to calculate the differential
transverse energy distribution of a jet in the CDF central rapidity strip,
$0.1<\left|  \eta\right|  <0.7$, while requiring that a second jet with
transverse energy at least as large as one half of the transverse energy of
the central jet, $E_{T,2}\geq E_{T,1}/2$, is tagged in a forward region,
$1.2<\left|  \eta_{2}\right|  <1.6$. \ The corresponding di-JEF probability
density function for an event obeying the appropriate cuts is expressed as
\begin{align}
P_{\mathrm{di-jet}}\left(  E_{T}\right)   &  =\frac{1}{\left(  \pi
R^{2}\right)  ^{2}}\iiiint d\eta_{1}d\phi_{1}d\eta_{2}d\phi_{2}\ \Omega\left(
0.1<\left|  \eta1\right|  <0.7\right)  \times\Omega\left(  1.2<\left|
\eta_{2}\right|  <1.6\right) \label{diJEF}\\
&  \times\Omega\left(  E_{T}\left(  J\left(  \eta_{2},\phi_{2}\right)
\right)  >E_{T}\left(  J\left(  \eta_{1},\phi1\right)  \right)  /2\right)
\times\delta\left(  E_{T}-E_{T}\left(  J\left(  \eta_{1},\phi_{1}\right)
\right)  \right)  .\nonumber
\end{align}
Using this probability function the desired differential cross section is
obtained from Eq. \ref{diffCn}%
\begin{equation}
\frac{d\sigma_{\mathrm{di-jet}}}{dE_{T}}=\frac{1}{\mathcal{L}}\sum
\limits_{\mathrm{events}}P_{\mathrm{di-jet}}\left(  E_{T}\right)  .
\label{diffdiJEF}%
\end{equation}
The perturbative results for this cross section at LO and NLO for both the
JEF\ analysis of Eq. \ref{diffdiJEF} and using the standard cone jet analysis
of EKS\cite{EKS} are exhibited in Figure 6 for the case $R=0.7$. \
%TCIMACRO{\FRAME{ftbpFU}{3.192in}{3.2638in}{0pt}{\Qcb{Differential cross
%section of Eq. \ref{diffdiJEF} for both JEF (moving cone) and EKS (fixed cone)
%analysis at LO and NLO.}}{\Qlb{fig4}}{fig4a.eps}%
%{\special{ language "Scientific Word";  type "GRAPHIC";
%maintain-aspect-ratio TRUE;  display "USEDEF";  valid_file "F";
%width 3.192in;  height 3.2638in;  depth 0pt;  original-width 6.2759in;
%original-height 6.4186in;  cropleft "0";  croptop "1";  cropright "1";
%cropbottom "0";  filename 'P5_ellis_0707_fig6.eps';file-properties "XNPEU";}}}%
%BeginExpansion
\begin{figure}
[ptb]
\begin{center}
\includegraphics[
height=3.2638in,
width=3.192in
]%
{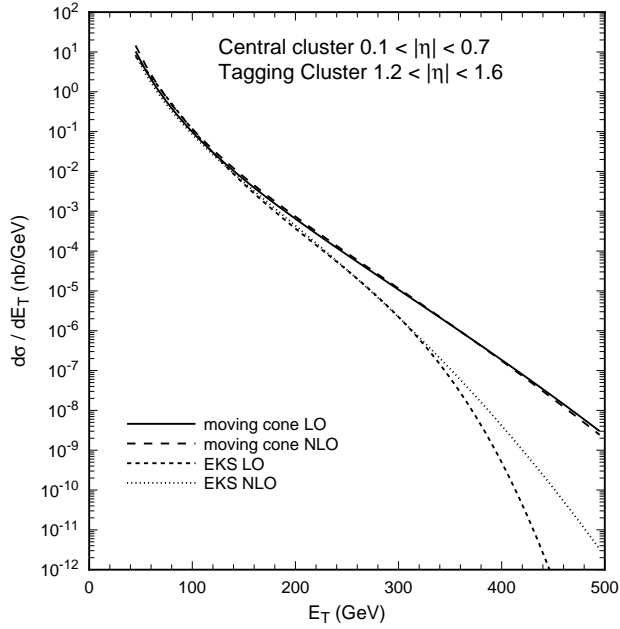}%
\caption{Differential cross section of Eq. \ref{diffdiJEF} for both JEF
(moving cone) and EKS (fixed cone) analysis at LO and NLO.}%
\label{fig4}%
\end{center}
\end{figure}
%EndExpansion
These results illustrate some of the desirable features of the JEF approach.
\ Note first that at smaller $E_{T}$ values both jet definitions yield similar
results at both LO and NLO. \ However, at larger $E_{T}$ values, where the
boundaries of phase space play are more relevant, the JEF result is larger
than the ``traditional'' cone jet result and, more importantly, is less
sensitive to the higher order corrections. \ We can understand this reduced
sensitivity in terms of the smearing of the rapidity cuts in the JEF analysis.
\ In the JEF analysis the underlying partons can violate the rapidity cuts by
as much as $R$ and still contribute to a JEF style jet that respects the cut.
\ For example, the parton contributing to the secondary jet is only required
to obey $\left|  \eta_{2}\right|  >1.2-R=0.5$ in order to make a nonzero
contribution. \ In contrast, the traditional cone jet with a single parton
inside requires the parton to be collinear with the jet and thus in the
current analysis receives contributions only from $\left|  \eta_{2}\right|
>1.2$. \ This smearing of the details of the rapidity cuts explains both the
larger magnitude and the reduced dependence on higher orders of the JEF di-jet
analysis. \ We can expect a similarly reduced dependence on the stochastic
effects of showering and hadronization.

\subsection{Concluding Remarks}

We have discussed a different approach to the jet analysis of hadronic states,
the JEF\ analysis, which follows from the earlier C-algebra 
formalism\cite{Tkachov}
and which differs from traditional algorithmic approaches in
that unique jets are not identified event-by-event. \ Instead the analysis
proceeds through the evaluation of Jet Energy Flow distributions. \ The brief
discussion presented here suggests that the JEF style analysis of hadronic
final states in hard scattering processes will provide observable measures of
the underlying short-distance parton structure that are less sensitive to
higher order corrections and to showering/hadronization corrections than more
conventional jet algorithm analyses. \ Clearly much more needs to be done in
order to demonstrate and make use of this conclusion.

\end{document}